\documentclass[superscriptaddress,aps,showpacs,nofootinbib,twocolumn]{revtex4}
\usepackage{graphicx}
\usepackage{amsmath,amssymb}

\begin{document}

\title{Braneworld inflation from an effective field theory after WMAP three-year data}

\author{M. C. Bento}
\email{bento@sirius.ist.utl.pt}
\affiliation{%
Departamento de F\'{\i}sica and Centro de F\'{\i}sica Te\'orica de
Part\'{\i}culas, Instituto Superior T\'{e}cnico, Avenida Rovisco
Pais, 1049-001 Lisboa, Portugal}

\author{R. Gonz\'{a}lez Felipe}
\email{gonzalez@cftp.ist.utl.pt}
\affiliation{%
Departamento de F\'{\i}sica and Centro de F\'{\i}sica Te\'orica de
Part\'{\i}culas, Instituto Superior T\'{e}cnico, Avenida Rovisco
Pais, 1049-001 Lisboa, Portugal}

\author{N.~M.~C. Santos}
\email{n.santos@thphys.uni-heidelberg.de}
\affiliation{%
Institut f\"{u}r Theoretische Physik, Universit\"{a}t Heidelberg\\
Philosophenweg 16, 69120 Heidelberg, Germany}

\begin{abstract}
In light of the results from the WMAP three-year sky survey, we
study an inflationary model based on a single-field polynomial
potential, with up to quartic terms in the inflaton field. Our
analysis is performed in the context of the Randall-Sundrum II
braneworld theory, and we consider both the high-energy and
low-energy (i.e. the standard cosmology case)  limits of the
theory. We examine the parameter space of the model, which leads
to both large-field and small-field inflationary type solutions.
We conclude that small-field inflation, for a potential with a
negative mass square term, is in general favored by current bounds
on the tensor-to-scalar perturbation ratio $r_s$.

\end{abstract}

\date{\today} \maketitle

\section{Introduction}

Inflation, originally introduced to solve the initial condition
problems of standard cosmology (SC)~\cite{Guth:1980zm_plus}, is at
present the favorite paradigm for explaining both the Cosmic
Microwave Background Radiation (CMBR) temperature anisotropies and
the initial conditions for structure formation. Indeed, it is
during this epoch of accelerated expansion that the primordial
density perturbations, which act as seeds for large-structure
formation in the universe, are generated through the amplification
of quantum fluctuations of the
field(s)~\cite{Mukhanov:1981xt_plus}. In the simplest inflationary
models, the energy density of the universe is dominated by a
single scalar field $\phi$, the so-called inflaton, that slowly
rolls down its self-interaction potential.

The recent publication of the three year results of the Wilkinson
Microwave Anisotropy Probe
(WMAP)~\cite{Hinshaw:2006ia_plus,Spergel:2006hy} continues to
support the standard inflationary predictions. WMAP data provides
no indication of any significant deviations from gaussianity and
adiabaticity of the CMBR power spectrum and suggests that the
universe is spatially flat to within the limits of observational
accuracy. Moreover, it allows for very accurate constraints on the
spectral index; the WMAP three-year data (WMAP3)
yield~\cite{Spergel:2006hy}
\begin{align}
n_s = 0.951^{+0.015}_{-0.019}~,
\end{align}
at $68\%$ confidence level (CL), for vanishing running and no
tensor modes. This result is in agreement with the one previously
obtained in Ref.~\cite{Sanchez:2005pi}, $n_s = 0.954 \pm 0.023$,
using  a joint analysis of the power spectrum of galaxy clustering
measured from the final two-degree field galaxy redshift survey
(2dfGRS) and a pre-WMAP3 compilation of measurements of both the
temperature power spectrum and the temperature-polarization
cross-correlation of the CMBR. A remarkable feature of these
results is that the Harrison-Zel'dovich scale-invariant spectrum
seems to be ruled out at around $3\sigma$ level\footnote{See
Ref.~\cite{Kinney:2006qm} for an analysis where the
scale-invariant spectrum is consistent with WMAP3 data.}.

Another feature of WMAP3 data is the evidence for a significant running of the
scalar spectral index~\cite{Spergel:2006hy},
\begin{align}
\alpha_s = -0.102^{+0.050}_{-0.043}~,
\end{align}
at $68\%$ CL and considering tensor perturbations\footnote{If
tensor perturbations are not taken into account slightly less
negative values are obtained.}. This evidence was already present
in the WMAP1 analysis \cite{Spergel:2003cb_plus} and persists when
large scale structure data is included~\cite{Spergel:2006hy},
although it is diluted by the addition of Lyman-$\alpha$ forest
data~\cite{Seljak:2004xh_plus,Lewis:2006ma,Seljak:2006bg}. We
should note however that zero running is at about $2\sigma$ from
the central value and hence WMAP3 does not demand a nonvanishing
running. In fact, vanishing running is still a good fit to the
CMBR data, and the improvement of the $\chi^2$ is small ($\Delta
\chi^2 = -3$), if running is allowed. Notice also that if both
tensor modes and running are taken into account, the WMAP team
obtained $n_s= 1.21^{+0.13}_{-0.16}$ as the best fit value for the
scalar tilt.

On the theoretical side, motivated by developments in
string/M-theory, there has been considerable interest in the
so-called braneworld constructions, where the matter fields are
trapped in a lower dimensional brane, while (in the simplest
models) only gravity can propagate into the bulk. Of special
interest for cosmology is the Randall-Sundrum II (RSII)
construction~\cite{Randall:1999vf}, consisting of a single brane
with positive tension embedded in a 5-dimensional bulk with a
negative cosmological constant (anti-de Sitter spacetime). A
remarkable feature of the RSII brane cosmology (BC) is the
modification of the expansion rate of the universe, $H=\dot{a}/a$,
before the nucleosynthesis era~\cite{Binetruy:1999hy_plus}. While
in standard cosmology the expansion rate scales with the energy
density $\rho$ as $H\propto\sqrt{\rho}$, this dependence becomes
$H\propto\rho$ at very high energies in brane cosmology. This
behavior may have important consequences on early universe
phenomena such as
inflation~\cite{Maartens:1999hf,Bento:2001hu_plus} and the
generation of the baryon asymmetry~\cite{Mazumdar:2000gj_plus}.

In this paper we study the RSII braneworld inflationary period in the light of
WMAP3 results, for a single-field power-law inflaton potential of the
form~\cite{Hodges:1989dw}
\begin{align}
V(\phi)= V_0 + \dfrac{1}{2}\, s\, m^2\, \phi^2 + \dfrac{1}{3} \, m\,
\widetilde{g} \,\phi^3 + \dfrac{1}{4}\, \widetilde{\lambda}\,
\phi^4~.\label{eq:potphi}
\end{align}
Here $s=1$ (unbroken symmetry) or $s=-1$ (broken symmetry) and $m>0$. The
parameters $\widetilde{\lambda}$ and $\widetilde{g}$ are dimensionless
constants; $\widetilde{\lambda}$ must be positive in order to ensure the
stability of the potential, while $\widetilde{g}$ can have either sign.

This potential has been studied previously in the literature in
the SC
context~\cite{Hodges:1989dw,Cirigliano:2004yh,Boyanovsky:2005pw,deVega:2006hb}
and covers a wide class of inflationary scenarios: for both
spontaneously broken and unbroken (negative and positive mass
square term, respectively) potentials one can obtain small and
large-field inflation (hereafter also called new and chaotic
inflationary solutions, respectively). Monomial potentials have
already been widely studied both in standard cosmology (for a
recent update following WMAP3 data see Refs.
\cite{Kinney:2006qm,Alabidi:2006qa_plus}) and in the braneworld
context~\cite{Maartens:1999hf,Bento:2001hu_plus,Tsujikawa:2003zd_plus};
these can be regarded as particular cases of the one considered
here. It is also worth noticing that potentials of the type we are
considering, Eq.~(\ref{eq:potphi}), can be obtained as
 effective potentials in some supergravity models (for a recent study see
Ref.~\cite{Ibe:2006fs}).

The paper is organized as follows. After discussing in
Section~\ref{sec:potential} the main features of the inflaton potential, we
briefly review in Section~\ref{sec:BraneInflation} the slow-roll inflation
formalism in the braneworld context. In Section~\ref{sec:results} we present
and discuss our results. Some final comments and conclusions are given in
Section~\ref{sec:conclusion}.

 \begin{figure*}[t]
\includegraphics[width=5.9cm]{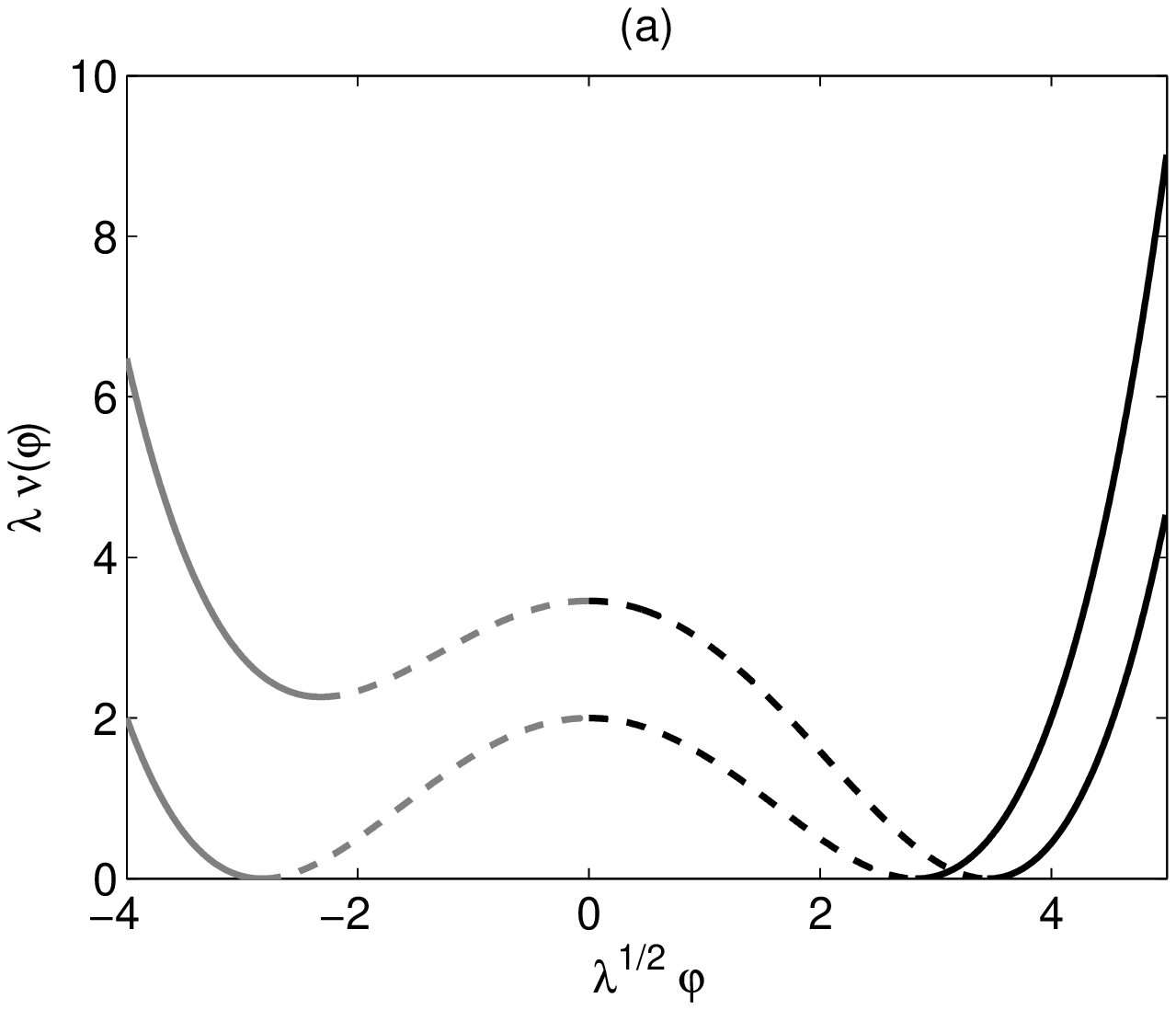}
\includegraphics[width=5.9cm]{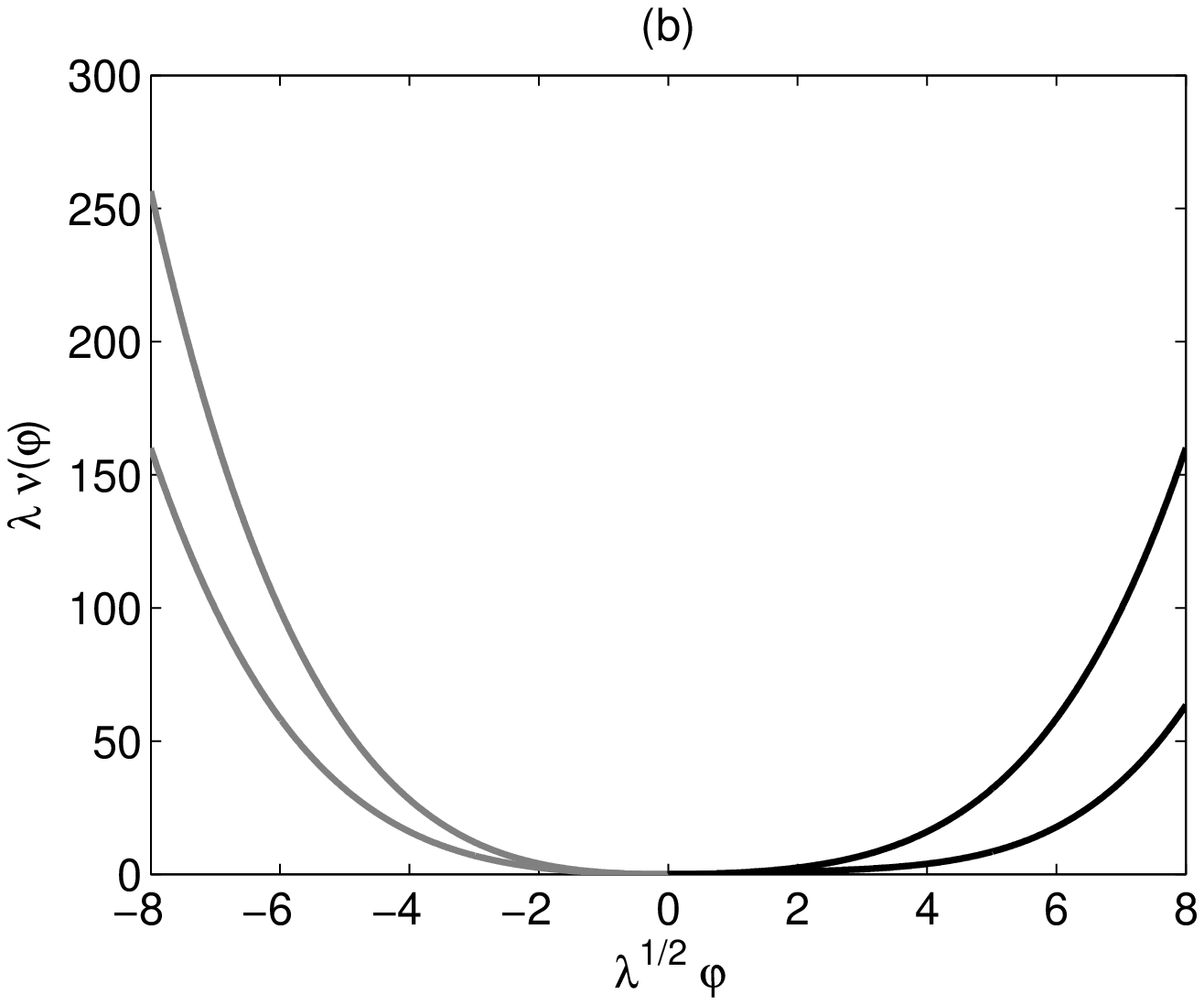}
\includegraphics[width=5.9cm]{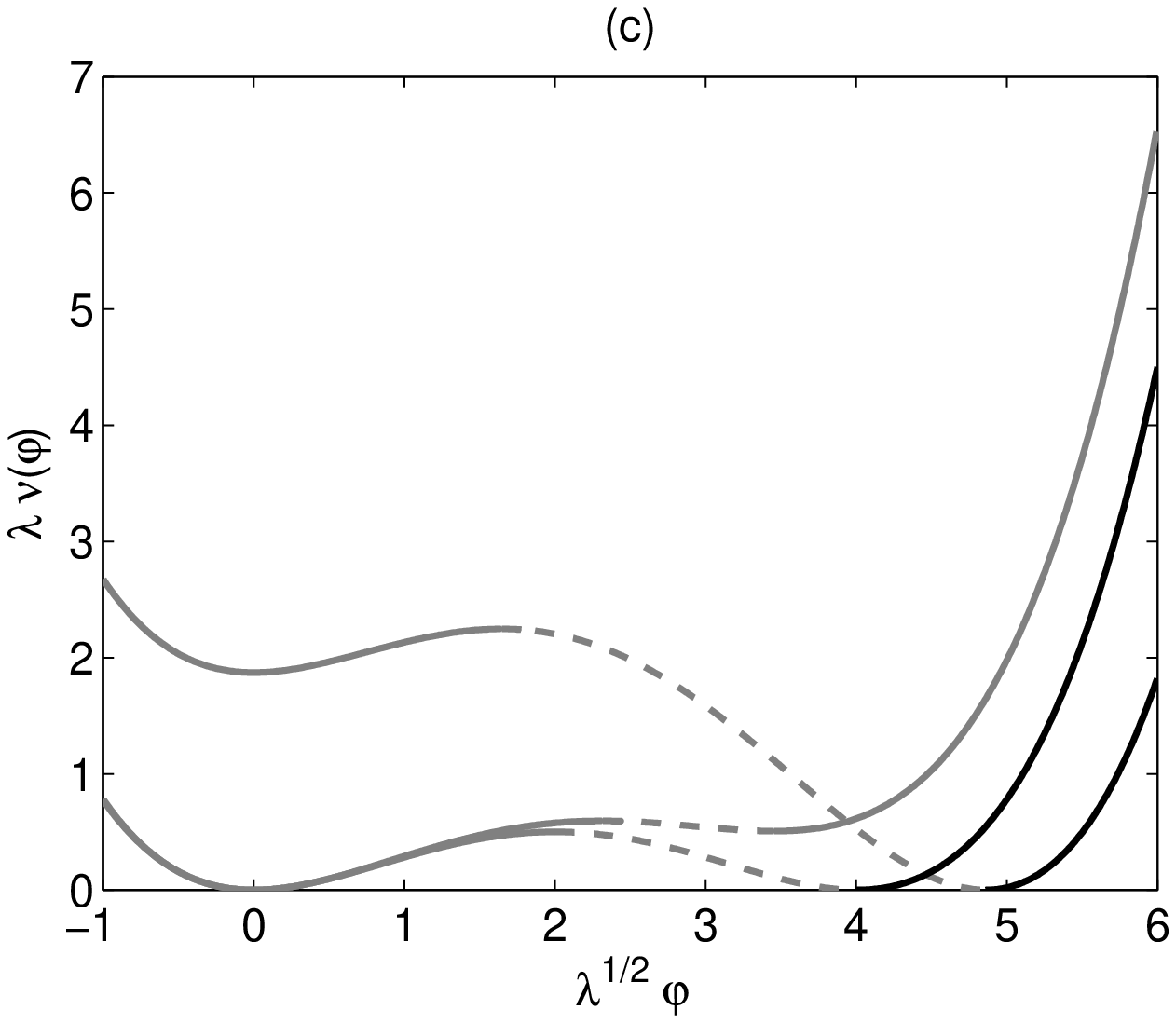}\\
\caption{The potential of Eq.~(\ref{eq:pot_dless}) for the cases discussed in
the text. The full (dashed) lines correspond to large (small) field inflation.
The curves correspond to different values of the asymmetry parameter $h$: (a)
$h=-0.2,0$, (b) $h=-0.8,0$ and (c) $h=-1.15,-\sqrt{9/8},-1.02$. We will only
consider inflation occurring on the black branches of the potential (cf.
Table~\ref{table:pot}).} \label{fig:potential}
\end{figure*}
\begin{table*}[t!]
\begin{tabular}{c c c c c c}
\hline \hline
&s & $h$ range & $\varphi$ range & Inflation type & Case\\
\hline
&$-1$ & $h \leq 0$ & $0 \lesssim \varphi < \varphi_+$ & Small-field (a)&  A\\
&$-1$ & $h \leq 0$ & $\varphi > \varphi_+$ & Large-field (a)&  B\\
&$+1$ & $-1 \leq h \leq 0$ & $\varphi > 0$ & Large-field (b)& C\\
&$+1$ & $-\sqrt{9/8} \leq h \leq -1 $ & $0 < \varphi \lesssim \varphi_-$ & Large-field (c)& D\\
&$+1$ & $h \leq -\sqrt{9/8}$ & $\varphi_- \lesssim \varphi < \varphi_+$ & Small-field (c)& E\\
&$+1$ & $h \leq -\sqrt{9/8}$ & $\varphi > \varphi_+$ & Large-field (c)& F\\
\hline \hline
\end{tabular}
\caption{Cases A-F correspond to different ranges of the potential
parameters and field values, see Eqs. (\ref{eq:pot_dless}) and
(\ref{eq:hdef}), leading to large and small-field inflation.
Labels (a)-(c) correspond to the cases shown in
Fig.~\ref{fig:potential}.} \label{table:pot}
\end{table*}

\section{Inflation from an effective field theory}
\label{sec:potential}

The potential of Eq.~(\ref{eq:potphi}) is the most general
renormalizable (power law) potential. Of course, more complicated
single-field potentials can be obtained from low-energy effective
field theories, but the terms in the potential of
Eq.~(\ref{eq:potphi}) can, in many cases, be regarded as the
leading terms in a power series expansion of such potentials, as
e.g. it is the case of supergravity models. Moreover, as noticed
in the Introduction, this single-field potential already covers
different types of inflationary scenarios, namely small and
large-field inflation.

In view of the stringent limits on the present vacuum energy density, a
reasonable assumption is to set the global minimum of the potential to zero.
This ensures that inflation does not run forever and ends with a finite number
of e-folds. This also fixes the parameter $V_0$ in terms of the potential
parameters $m$, $\widetilde{g}$ and $\widetilde{\lambda}$. One should also
notice that the potential has a $\phi \rightarrow -\phi$, $\widetilde{g}
\rightarrow -\widetilde{g}$ symmetry. Thus, in order to explore the parameter
space it is sufficient to choose a definite sign for $\widetilde{g}$. In our
analysis we will consider $\widetilde{g} \leq 0$ and assume $\phi$ to be
positive.

In order to analyze the potential it is useful to redefine the inflaton field
as
\begin{align}
\varphi \equiv \dfrac{\phi}{M_P}~,
\end{align}
where $M_P \simeq 2.4 \times 10^{18}$~GeV is the reduced
four-dimensional Planck mass, and rewrite the potential as
\begin{align}
V(\varphi)= m^2\, M_P^2 \;v(\varphi)~,\label{eq:potvarphi}
\end{align}
where the dimensionless part is given by
\begin{align}
v(\varphi)=v_0 + \dfrac{1}{2}\, s\, \varphi^2 + \dfrac{2}{3} \,g \,\varphi^3 + \dfrac{1}{32}\,
\lambda\, \varphi^4~,\label{eq:pot_dless}
\end{align}
with
\begin{align}
v_0=\dfrac{V_0}{m^2\, M_P^2}~,~ \quad g=\dfrac{\widetilde{g}\, M_P}{2\, m}~,~
\quad \lambda=\dfrac{8 \,\widetilde{\lambda}\, M_P^2}{m^2}~.
\end{align}

The qualitative behavior of the potential is easily understood with the
determination of the critical points (maxima, minima or inflection points),
obtained from $v'(\varphi)=0$, where the prime denotes the derivative with
respect to the dimensionless field $\varphi$. There is a critical point at
$\varphi_0=0$ and two more possible critical points (depending on the
parameters $g$ and $\lambda$) at
\begin{align}
\varphi_{\pm} = \dfrac{ 8\,|g| \pm 2
\sqrt{16\,g^2-2\,s\,\lambda}}{\lambda}~,
\end{align}
where we are restricting to the case $g \leq 0$.

It is easy to see that for $s=-1$ the potential has one maximum at
$\varphi_0=0$ and a global minimum at $\varphi_+$ (there is also a
local minimum at $\varphi_-$, which for $g=0$ is degenerated with
the first one) - see Fig.~\ref{fig:potential}(a). The parameter
$g$ determines the asymmetry of the potential: the larger its
absolute value is, the more asymmetric is the potential. In this
case, one can have both large and small-field inflation, depending
on the initial value of the inflaton field (cases A and B in Table
I).

For $s=1$ we can distinguish two cases. If $\lambda \geq 8 g^2$,
the only critical point of the potential is $\varphi_0=0$,
corresponding to the minimum of the potential. Once more, the
larger $|g|$ is, the more asymmetric the potential becomes [see
Fig.~\ref{fig:potential}(b)]. In this case only large-field
inflation is possible. If $\lambda < 8 g^2$,  the potential has
three critical points: one maximum at $\varphi_-$ and two minima
at $\varphi_0$ and $\varphi_+$. For $\lambda= 64\, g^2/9$ the
minima are degenerated, for $\lambda> 64\, g^2/9$ the global
minimum is at $\varphi_0$ and for $\lambda< 64\, g^2/9$ at
$\varphi_+$ [see Fig.~\ref{fig:potential}(c)]. In this parameter
range we can have either small or large-field inflation.

It is useful to reformulate the previous discussion in terms of the parameter
\begin{align}
h=g\,\sqrt{\dfrac{8}{\lambda}}~. \label{eq:hdef}
\end{align}
For $s=1$ and $-1 \leq h \leq 0$ (we are considering only negative
values of $g$), the potential has only one minimum at
$\varphi_0=0$. In this case only large-field inflation is
possible, corresponding to case C of Table \ref{table:pot}. For $h
< -\sqrt{9/8}$, the global minimum of the potential is at
$\varphi_+$ (and there is a local minimum at $\varphi_0=0$). Both
large and small-field inflation are possible but in order to
obtain small-field inflation, case E of Table \ref{table:pot}, the
initial value of the field should lie between $\varphi_-$ and
$\varphi_+$, thus implying a certain amount of fine-tuning; hence,
we will consider only the large-field case. For the same reason,
we will neglect the range $-\sqrt{9/8}<h<-1$, where the global
minimum of the potential is at $\varphi_0=0$ and there is also a
local minimum at $\varphi_+$, case D of Table \ref{table:pot}.
Notice also that, in Table~\ref{table:pot},  only the cases where
the field rolls towards the global minimum are taken into
consideration.

\begin{table*}[ht!]
\begin{tabular}{c c c c c}
\hline\hline & $n_s^{high}$ & $r_s^{high}$ & $n_s^{low}$ &
$r_s^{low}$\\\hline\\
\hspace{5mm}$p$\hspace{5mm}
&\hspace{4mm}$\dfrac{(p+2)\,N_\star-3\,p-2}{(p+2)\,N_\star+p}$\hspace{4mm}
&\hspace{4mm}$\dfrac{24\,p}{(p+2)\,N_\star+p}$\hspace{4mm}&
\hspace{4mm}$\dfrac{2\,N_\star-3}{2\,N_\star-1+p}$\hspace{4mm}&
\hspace{4mm}$\dfrac{8\,p}{2\,N_\star-1+p}$\hspace{4mm}\\\\
$2$ & $\dfrac{2\,(N_\star-2)}{2\,N_\star+1}=0.959$ &
$\dfrac{24}{2\,N_\star+1}=0.20$ & $\dfrac{2\,N_\star-3}{2\,N_\star+1}=0.967$
& \hspace{4mm}$\dfrac{16}{2\,N_\star+1}=0.13$\\\\
$4$ & \hspace{4mm}$\dfrac{3\,N_\star-7}{3\,N_\star+2}=0.951$\hspace{4mm}&
\hspace{4mm}$\dfrac{48}{3\,N_\star+2}=0.26$\hspace{4mm} &
 \hspace{4mm}$\dfrac{2\,N_\star-3}{2\,N_\star+3}=0.951$\hspace{4mm}
& \hspace{4mm}$\dfrac{32}{2\,N_\star+3}=0.26$\hspace{4mm}\\\\
\hline\hline
\end{tabular}
\caption{High- and low-energy limits in the RS II braneworld model for the
scalar spectral index $n_s$ and the tensor-to-scalar perturbation ratio $r_s$,
in the case of a simple monomial inflationary potential of the form $V \propto
\phi^p$. The numerical values are computed taking $N_\star=60$.}
\label{table:nsrs}
\end{table*}

\section{Slow-roll braneworld inflation}
\label{sec:BraneInflation}

We will consider the RSII scenario, where all matter fields are
confined to the brane and, hence, inflation is driven by a 4D
scalar field trapped on the brane. In this scenario, the detailed
form of perturbations produced by the inflationary potential is
modified due to the modification of the Friedmann equation at high
energies and because gravitational wave perturbations are able to
go into the bulk. In fact, in a cosmological scenario in which the
metric projected onto the brane is a spatially flat
Friedmann-Lema\^{\i}tre-Robertson-Walker model, the Friedmann
equation in 4D acquires an extra term,
becoming~\cite{Binetruy:1999hy_plus}
\begin{align}
H^2 = {1 \over 3\, M_P^2}\, \rho\, \left[1 + {\rho \over 2 \widetilde{\sigma}}\right]~,
\label{eq:Friedmann}
\end{align}
after setting the four-dimensional cosmological constant to zero and assuming
that inflation rapidly makes any dark radiation term negligible. The brane
tension $\widetilde{\sigma}$ relates the four- and five-dimensional Planck
masses through the relation
\begin{align}
\widetilde{\sigma} = \frac{3}{32 \pi^2} \frac{M_5^6}{M_P^2}~,
\label{eq:tension}
\end{align}
where $M_5$ is the 5D Planck mass. Notice that
Eq.~(\ref{eq:Friedmann}) reduces to the usual Friedmann equation
at sufficiently low energies, \mbox{$\rho \ll
\widetilde{\sigma}$}, while at very high energies we have
$H\propto\rho$. Successful big bang nucleosynthesis (BBN) requires
that the change in the expansion rate due to the new terms in the
Friedmann equation be sufficiently small at scales $\sim
\mathcal{O}$(MeV); this implies $M_5 \gtrsim
40~\mbox{TeV}$~\cite{Cline:1999ts_plus}. A more stringent bound,
$M_5 \gtrsim 10^5~\mbox{TeV}$, is obtained by requiring the theory
to reduce to Newtonian gravity on scales larger than
1~mm~\cite{Maartens:1999hf}. For the sake of convenience, in the
following we shall work with a dimensionless brane tension defined
as
\begin{align}
{\sigma} = \frac{\widetilde{\sigma}}{m^2\, M_P^2}~. \label{eq:tension_dless}
\end{align}

Since the scalar field is confined to the brane, the conservation equation
implies that the field satisfies the usual Klein-Gordon equation,
\begin{align}
\ddot \varphi + 3 \, H\, \dot \varphi + m^2\, v'(\varphi) = 0~,
\label{eq:KG}
\end{align}
where $v(\varphi)$ is the dimensionless potential defined in
Eq.~(\ref{eq:pot_dless}), and the prime denotes the derivative
with respect to the dimensionless field $\varphi$. The high-energy
corrections provide increased Hubble damping,
\begin{align}
H^2 \simeq {m^2 \over 3} \,v \,\left[1 + {v \over 2\, \sigma}\right]~,
\label{eq:Friedmann_pot}
\end{align}
which makes the evolution of the inflaton slower.

\begin{figure*}[t]
\includegraphics[width=14cm]{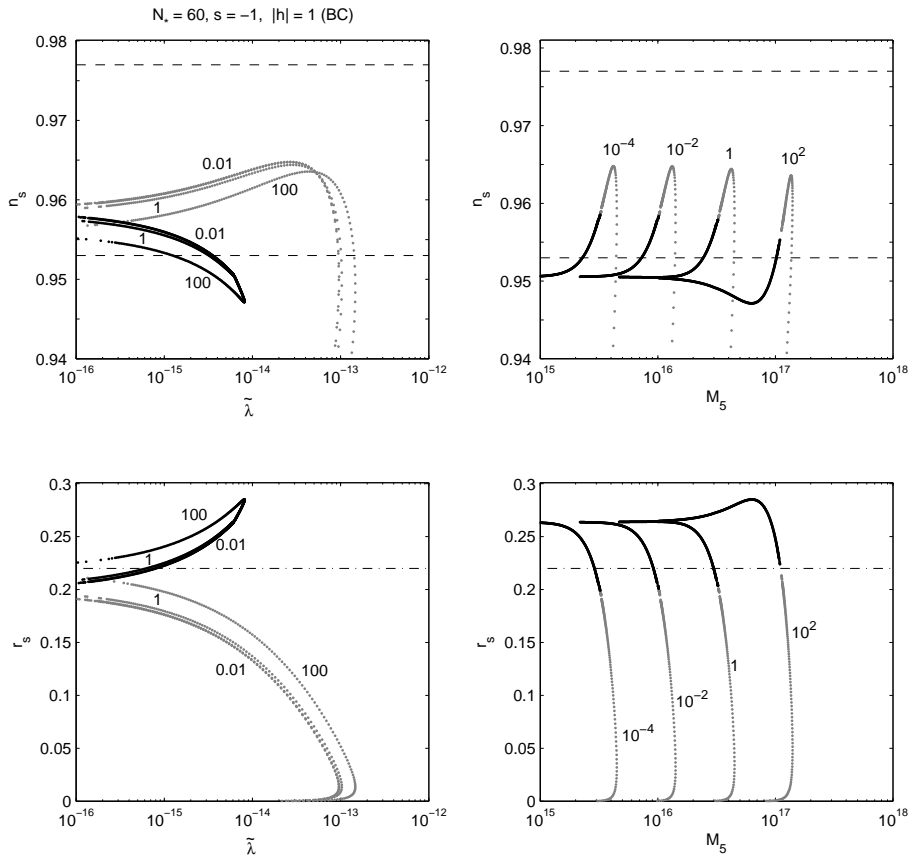}\\
\caption{$n_s$ as a function of $\widetilde{\lambda}$ and $M_5$ (upper panel)
and $r_s$ as a function of $\widetilde{\lambda}$ and $M_5$ (lower panel), in
the RSII braneworld model, for the potential of Eq.~(\ref{eq:potphi}), broken
symmetry case.  Gray (black) lines indicate small (large) field inflationary
solutions (respectively, cases A and B of Table I). The numbers $10^{-4}-10^2$
refer the value of the brane tension $\sigma$. The asymmetry parameter h is
fixed at $\vert h\vert=1$ and we have taken $N_\star=60$. The horizontal dashed
lines indicate the observational bounds on $n_s$ and the dot-dashed lines
correspond to the upper bound on $r_s$, see Eqs. (\ref{eq:nsbound}) and
(\ref{eq:rsbound}).}
 \label{fig:bcbroken}
\end{figure*}
\begin{figure*}[t]
\includegraphics[width=14cm]{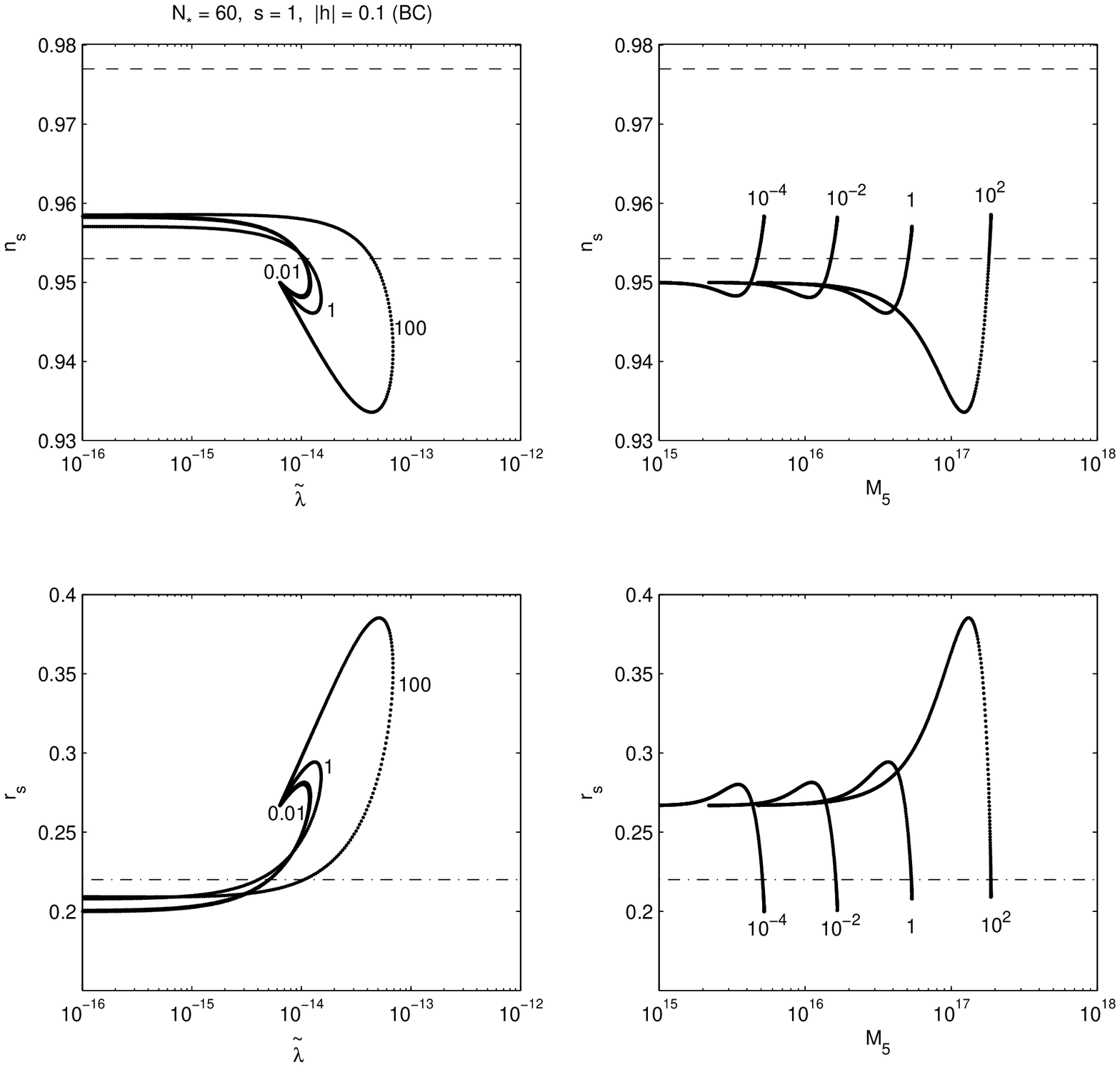}\\
\caption{As in Fig.~\ref{fig:bcbroken}, but for the case of
unbroken symmetry and large-field inflation (case C of Table I).
We take $\vert h\vert = 0.1$.} \label{fig:bcunbroken}
\end{figure*}

The number of e-folds during the inflationary period is given in the slow-roll
approximation by \cite{Maartens:1999hf}
\begin{align}
N(\varphi) \simeq -  \int_{\varphi}^{\varphi_{\rm F}}\dfrac{ v }{ v'} \left[
1+{v \over 2\,\sigma}\right] d{\varphi}~,\label{eq:Nfolds}
\end{align}
where the subscript F corresponds to the end of inflation. Braneworld effects
at high energies increase the Hubble rate by a factor $v/2\,\sigma$, yielding
more inflation between any two values of $\varphi$ for a given potential. The
value of $\varphi$ at the end of inflation can be obtained from the condition
\begin{align}
{\rm max}\{\epsilon(\varphi_F),|\eta(\varphi_F)|\}= 1~,
\label{eq:phif}
\end{align}
where the slow-roll parameters are now defined as
\begin{align}
\epsilon & = {1 \over 2} \left( {v' \over v}\right)^2 {1+{v/ \sigma}\over(1+{v/ 2
\,\sigma})^2}~,
\label{eq:epsilon}\\
\eta & =  {v'' \over v}   {1 \over 1+{v/ 2\, \sigma}}~. \label{eq:eta}
\end{align}

The prediction for the inflationary variables typically depends on
the number of e-folds of inflation occurring after the observable
universe leaves the horizon, $N_\star = N(\varphi_\star)$.
Although a wide variety of assumptions about $N_\star$ can be
found in the literature, the determination of this quantity
requires a model of the entire history of the universe. However,
while from nucleosynthesis onwards this is now well established,
at earlier epochs there are considerable uncertainties such as the
mechanism ending inflation and details of the reheating process.
This issue has been recently reviewed in
Refs.~\cite{Liddle:2003as,Dodelson:2003vq}, where a
model-independent upper bound was derived, namely $N_\star<60$. In
fact, $N_\star=55$ is found to be a reasonable fiducial value with
an uncertainty of about 5 e-folds around that value. However, the
authors stress that there are several ways in which $N_\star$
could lie outside that range, in either direction. Moreover, in
the braneworld context, one expects $N_\star$ to depend on the
brane tension. Actually, larger values of $N_\star$ are expected
because, in the high-energy regime, the expansion laws
corresponding to matter and radiation domination are slower than
in the standard cosmology, which implies a greater change in $a H$
relative to the change in $a$, therefore  requiring a larger value
of $N_\star\,$. This was confirmed by the results of
Ref.~\cite{Wang:2003qr}, where the bound $N_\star < 75$ was found
for brane-inspired cosmology. In what follows we shall use
$N_\star = 60$.

In the RSII model, the scalar and tensor perturbation amplitudes are given
by~\cite{Maartens:1999hf,Langlois:2000ns}
\begin{align}
A_s^2 &= \dfrac{m^2}{75 \,\pi^2\, M_P^2}\,\dfrac{v^3}
{v^{\prime2}} \left[ 1 + \dfrac{v}{2\,\sigma} \right]^3 ~, \label{eq:As}\\
A_t^2 &=\dfrac{m^2}{150 \,\pi^2\, M_P^2}\,  v \, \left( 1+\dfrac {v}{ 2\, \sigma}\right)
F^2~, \label{eq:At}
\end{align}
where
\begin{align} F^2=\left[\sqrt{1+x^2} -x^2
\sinh^{-1}\left({1\over x}\right)\right]^{-1}~,\label{eq:AtF}
\end{align}
and
\begin{align}
x\equiv \left(\dfrac{3 \,H^2}{4 \pi\, m^2\,\sigma}\right)^{1/2} =
\left[\dfrac{2\,v} {\sigma
}\left(1+\dfrac{v}{2\,\sigma}\right)\right]^{1/2}~.\label{eq:Ats}
\end{align}
In the low-energy limit ($x\ll 1$), $F^2\approx 1$, whereas $F^2\approx 3\,
v/2\,\sigma$ in the high-energy limit. The right hand side of
Eqs.~(\ref{eq:As}) and ~(\ref{eq:At}) should be evaluated at horizon crossing,
$k=a H$, where $k$ is the commoving wavenumber, which in terms of the inflaton
field, corresponds to setting  $\varphi=\varphi_\star$.

The mass parameter $m$ can be determined from Eq.~(\ref{eq:As}),
\begin{align}
m= 5 \sqrt{3} \,\pi\,M_P\,  \dfrac{v^{\prime }} {v^{3/2}} \left[ 1 +
\dfrac{v}{2\,\sigma} \right]^{-3/2}A_s^{cmb}~, \label{eq:mass}
\end{align}
where $A_s^{cmb}$ is given by the WMAP amplitude of density fluctuations $A_s^2
\approx 4.72 \times 10^{-10}A$ with $A(k=0.002~ {\rm Mpc}^{-1}) \sim 0.8$.
Notice that while in SC the mass parameter $m$ is fixed by the other potential
parameters, in BC there is an extra degree of freedom - the brane tension.

The tensor power spectrum can be parameterized in terms of the tensor-to-scalar
ratio as
\begin{align}
r_s\equiv 16 \,\dfrac{A_t^2}{ A_s^2}~,\label{eq:rs}
\end{align}
where we have chosen the normalization so as to be consistent with the one of
Refs.~\cite{Peiris:2003ff,Spergel:2006hy}, in the low-energy limit.
WMAP3~\cite{Spergel:2006hy} alone gives $r_s < 0.55$ (with vanishing running)
and $r_s < 1.5$ (with running), both at $95\%$ CL. However, models with higher
values of $r_s$ require larger values of $n_s$ and lower amplitude of the
scalar fluctuations in order to fit the CMBR data, and these are in conflict
with large scale structure measurements (in the case of vanishing
running\footnote{If running index is allowed the large tensor components are
consistent with the data.}). Hence the strongest overall constraints on the
tensor mode contribution comes from the combination of CMBR and large scale
structure data. The combination of WMAP3 and Sloan Digital Sky Survey (SDSS)
measurements~\cite{Spergel:2006hy} give: $r_s < 0.28$ (without~running) and
$r_s < 0.67$ (with running), at $95\%$ CL. If the Lyman-$\alpha$ forest
spectrum from SDSS is also considered, then~\cite{Seljak:2006bg} $r_s < 0.22$
(at $95\%$ CL and without running).

The spectral tilt for scalar perturbations can be written in terms of the
slow-roll parameters as ~\cite{Maartens:1999hf}
\begin{align}
n_{s} - 1  & \equiv {d\ln A_{s}^2 \over d\ln k} = -6\,\epsilon +
2\,\eta~. \label{eq:ns}
\end{align}
The tensor index
\begin{align}
n_t &\equiv {d\ln A_{t}^2 \over d\ln k}\label{eq:nt}
\end{align}
obeys a more complicated equation~\cite{Huey:2001ae}. However, it can be shown
that the consistency relation
\begin{align}
n_t=-\dfrac{r_s}{8}\label{eq:nt2}
\end{align}
holds independent of the brane tension and, consequently, it has precisely the
same form as the one obtained in standard cosmology. This means that
perturbations do not contain any extra information as compared to SC, and, in
particular, that they cannot be used to determine the brane tension: for any
value of $\sigma$ one can always find a potential that generates the observed
spectra~\cite{Liddle:2001yq}.

The running of the scalar spectral tilt
\begin{align}
\alpha_{s}  \equiv {d n_{s} \over d\ln k} \label{eq:alphas}
\end{align}
can also be written in terms of the slow-roll parameters for the two limiting
cases
\begin{align}
\alpha_{s} & = 16\,\epsilon\,\eta - 24\,\epsilon^2 - 2\,\xi
\quad\quad {\rm for}~~ v/\sigma \ll 1~~{\rm (SC)}~,\label{eq:alphasSC}\\
\alpha_{s} & = 16\,\epsilon\,\eta - 18\,\epsilon^2 - 2\,\xi
\quad\quad {\rm for}~~v/\sigma
\gg 1~~{\rm (BC)}~, \label{eq:alphasBC}
\end{align}
where
\begin{align}
\xi = \dfrac {v'\, v'''}{ v^2} \dfrac{1 }{ \left(1+{v/ 2\,\sigma}\right)^2}~
\label{eq:xi}
\end{align}
is the ``jerk" parameter.

\begin{figure*}[t]
\includegraphics[width=14cm]{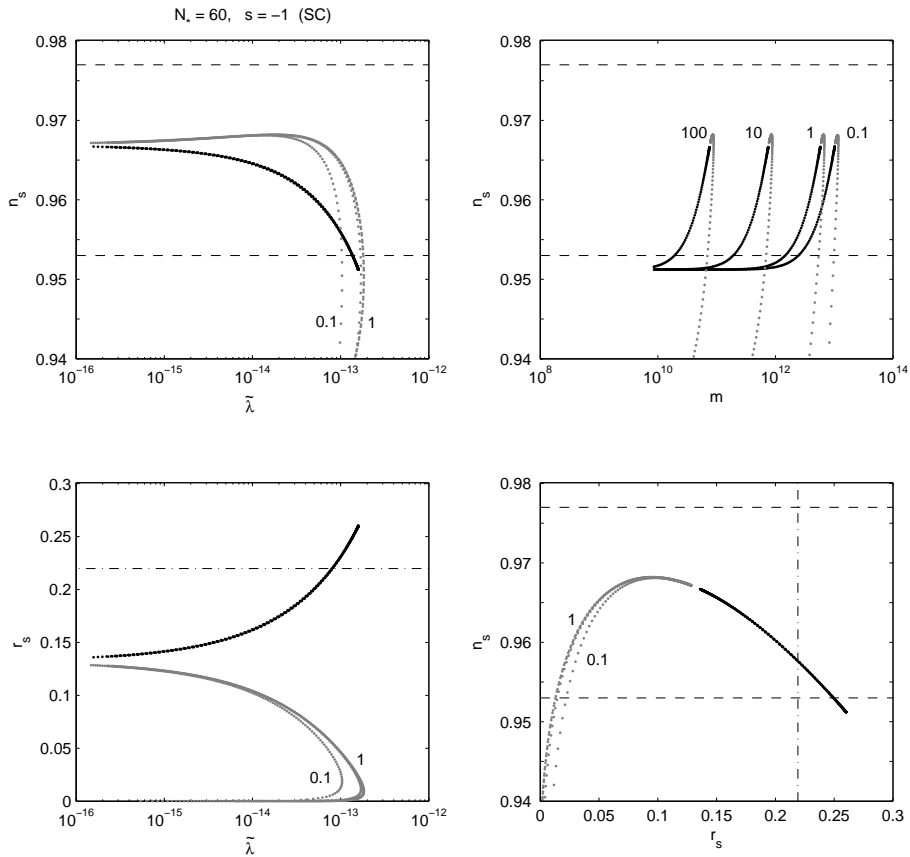}\\
\caption{$n_s$ as a function of $\widetilde{\lambda}$ and $m$ (upper panel) and
$r_s$ as a function of $\widetilde{\lambda}$ and $n_s$ (lower panel), in the SC
limit, for the potential of Eq.~(\ref{eq:potphi}), broken symmetry case. Gray
(black) lines indicate small (large) field inflationary solutions (cases A and
B of Table I, respectively). The numbers $0.1-100$ refer to the value of $\vert
h \vert$ for each curve and $N_\star=60$ is assumed. Horizontal dashed lines
indicate the observational bounds on $n_s$ and the dot-dashed lines correspond
to the upper bound on $r_s$, see Eqs. (\ref{eq:nsbound}) and
(\ref{eq:rsbound}).} \label{fig:scbroken}
\end{figure*}
\begin{figure*}[t]
\includegraphics[width=14cm]{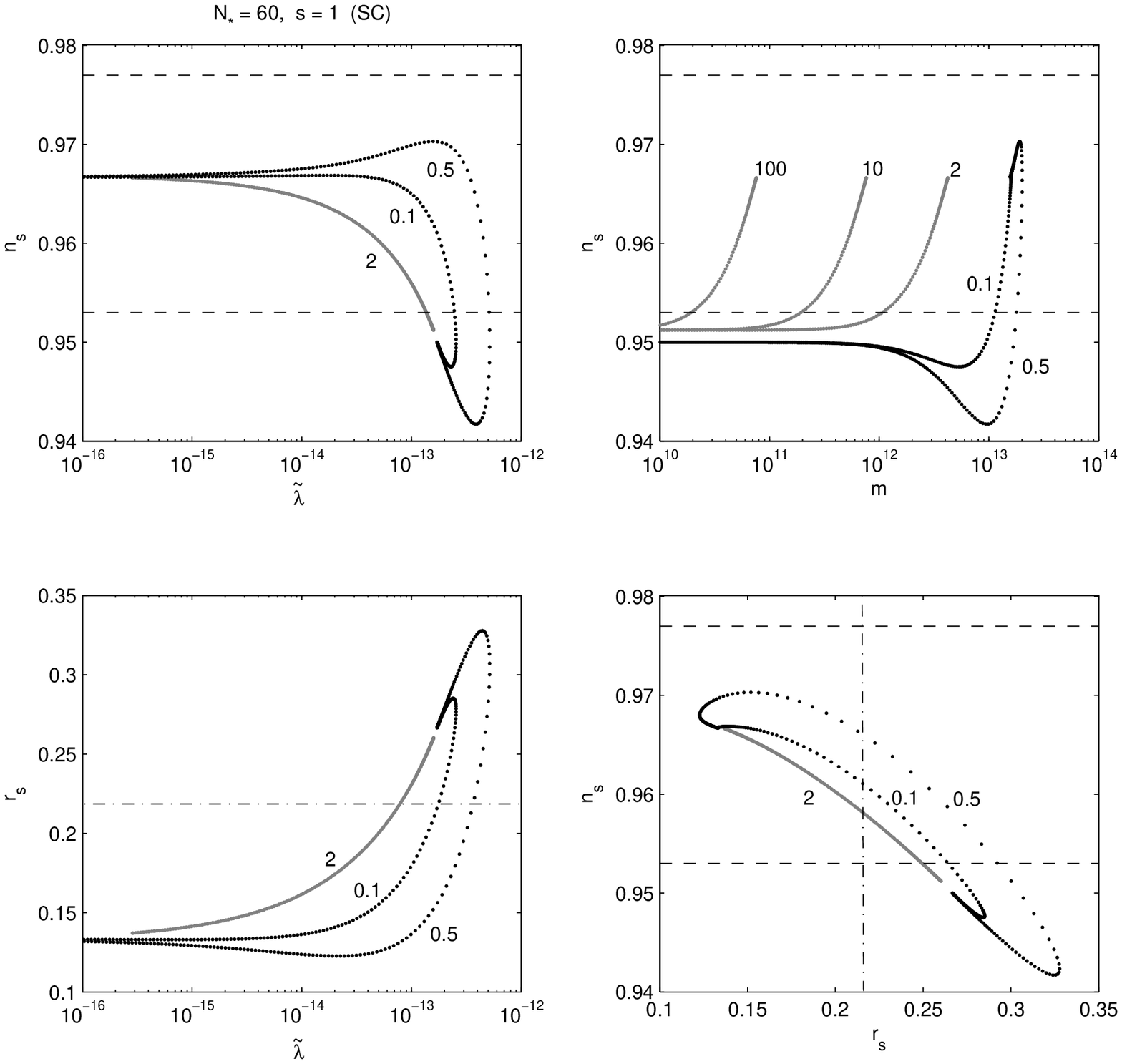}\\
\caption{As for Fig.~\ref{fig:scbroken} but for the case of
unbroken symmetry, large-field inflation (black and gray lines
correspond to cases C and F of Table I, respectively).}
\label{fig:scunbroken}
\end{figure*}

In the following we will analyze the different regimes (broken/unbroken,
small/large field) for  the potential of Eq. (\ref{eq:potvarphi}), in the
braneworld context,  taking into account the bounds obtained from the WMAP3
data. For completeness we will also study the low-energy limit ($v/ \sigma \ll
1$), i.e. the standard cosmology limit. It is also interesting to consider the
limiting cases of the monomial potentials $V \propto \phi^p$, for $p=2$ or $4$,
obtained, for $s=1$, when $\lambda \rightarrow 0$ and $\lambda \gg 1$,
respectively. In these cases the inflationary observables depend only on the
number of e-folds of inflation occurring after the observable universe leaves
the horizon, $N_\star$. In Table~\ref{table:nsrs} we show the values of the
scalar spectral index and ratio of tensor-to-scalar perturbations, as a
function of $N_\star$, for monomial potentials of the form $V \propto \phi^p$,
both in the high- and low-energy limits of brane cosmology.

We also remark in the models under consideration, the running $\alpha_s \sim
10^{-4}$ is very small, and therefore, one can make use of the observational
bounds obtained in the case of vanishing running. If besides the   WMAP3 data
we also take into account the galaxy clustering and supernovae data, as well as
the Lyman-$\alpha$ forest power spectrum data from SDSS, these bounds
are~\cite{Seljak:2006bg}
\begin{align}
n_s & = 0.965 \pm 0.012~ \left(\,^{+0.025}_{-0.024}\,\right)~,\label{eq:nsbound}\\
r_s & < 0.22~ \left(\,<0.37\,\right)~.\label{eq:rsbound}
\end{align}
The error bars are at $1\sigma$ ($2\sigma$) and the upper bounds
at $95\%$ ($99.9\%$) CL. The recent analysis of Kinney \textit{et
al}~\cite{Kinney:2006qm} is in good agreement with
Refs.~\cite{Lewis:2006ma,Seljak:2006bg}, but shows significant
inconsistencies with the results of WMAP3
team~\cite{Spergel:2006hy}. Moreover, in their analysis the
Harrison-Zel'dovich scale-invariant spectrum, with no running and
no tensor component, is consistent with the WMAP3 data alone at
95$\%$ CL. In our numerical analysis and the discussion that
follows we shall use the bounds given in Eqs.~(\ref{eq:nsbound})
and (\ref{eq:rsbound}).

\section{Results and discussion}
\label{sec:results}

In this Section we present the predictions for the density fluctuations in the
RSII braneworld context for the case of the inflationary potential given by
Eq.~(\ref{eq:potphi}).

In Fig.~\ref{fig:bcbroken} we plot $n_s$ as a function of $\widetilde{\lambda}$
and $M_5$ (upper panel) and $r_s$ as a function of $\widetilde{\lambda}$ and
$M_5$ (lower panel) for the broken symmetry case ($s=-1$). Gray (black) lines
indicate small (large) field inflationary solutions, which correspond to cases
A and B of Table I, respectively. For illustration, the asymmetry parameter
$\vert h \vert$ is fixed at $\vert h \vert =1$. The numbers $0.1-100$ refer to
the value of the brane tension $\sigma$ for each curve, and $N_\star=60$ is
assumed. Horizontal dashed lines indicate (upper and lower) observational
bounds on $n_s$ and dot-dashed lines the upper bound on $r_s$ as given in
Eqs.~(\ref{eq:nsbound}) and (\ref{eq:rsbound}).

We see from Fig.~\ref{fig:bcbroken} that, in the broken symmetry case, the new
inflationary solutions are favored as compared with chaotic ones when we take
into consideration the observational bound on $r_s$. In fact, chaotic
inflationary solutions for a quartic polynomial potential, yield larger values
of $r_s$ than the new inflationary ones for a given value of $n_s$. This
feature is also present in standard cosmology, as first pointed out in Ref.
\cite{Cirigliano:2004yh}. For comparison, the SC case is presented in
Fig.~\ref{fig:scbroken}, where we show $n_s$ as a function of the quartic
coupling $\widetilde{\lambda}$ and the mass parameter $m$ (upper panel) and
$r_s$ as a function of $\widetilde{\lambda}$ and $n_s$ (lower panel), for
different values of $|h|$. Notice also that the fact that new inflation is
favored over chaotic inflation is even more evident in the high-energy regime
of BC than in the SC limit. This is already apparent from Table II, for the
limiting case of the $\varphi^2$ potential, since we need higher values of
$N_\star$ to satisfy the $r_s$ bound in BC ($N_\star > 54$) than in SC
($N_\star > 36$). As expected, in order to reproduce the CMB density
fluctuations, the value of the quartic coupling $\widetilde{\lambda}$ should be
small; we get the  bound $\widetilde{\lambda} \lesssim 10^{-13}$.

In Fig.~\ref{fig:bcunbroken} we show the results for the case of unbroken
symmetry, choosing $|h|=0.1$, which corresponds to case C of Table I (case F
leads to results that are similar to case B, see Fig. \ref{fig:bcbroken}).
Remarkably, this type of solutions will be excluded for $r_s < 0.2$, noticeably
close to the bound given in Eq.~(\ref{eq:rsbound}). Comparing with the
corresponding case in SC (cf. Fig. \ref{fig:scunbroken}), we see that the
high-energy regime of BC is indeed much more constrained by the $r_s$ bound. In
this case, we get the upper bound $\widetilde{\lambda} \lesssim 10^{-12}$. This
bound, together with the relation
$|\widetilde{g}|=2\,|h|\,\widetilde{\lambda}^{1/2}$ and the condition $|h| \leq
1$, leads then to the following bound on the cubic coupling:  $|\widetilde{g}|
\lesssim 10^{-6}$.

Let us now briefly comment on the inflaton mass scale $m$, which is fixed by
the amplitude of the scalar adiabatic fluctuations, as can be readily seen from
Eq.~(\ref{eq:mass}). In the SC case and for small values of $\vert h \vert$,
using the e-fold equation (\ref{eq:Nfolds}) one expects $\varphi \sim
\mathcal{O}(\sqrt{N}),\, v(\varphi) \sim \mathcal{O}(N)$ and $v'(\varphi) \sim
\mathcal{O}(\sqrt{N})$ at horizon exit. Thus, we obtain
\begin{align}
m \simeq 5 \sqrt{3} \,\pi\,M_P\,\frac{A_s^{cmb}}{N_\star} \simeq 2 \times
10^{13}~\textrm{GeV}.
\end{align}
On the other hand, for large values of $\vert h \vert$, one can show that there
is an additional suppression factor $\mathcal{O}(1/|h|)$. In this case, higher
values of the asymmetry parameter $\vert h \vert$ would require smaller values
of the inflaton mass $m$ in order to correctly reproduce $A_s^{cmb}$. Such a
behavior is also evident from Figs.~\ref{fig:scbroken} and Fig.
\ref{fig:scunbroken}.

In the high-energy BC regime, Eq.~(\ref{eq:Nfolds}) suggests the scaling
behavior $\varphi \sim \mathcal{O}[(\sigma N)^{1/4}],\, v(\varphi) \sim
\mathcal{O}[(\sigma N)^{1/2}]$ and $v'(\varphi) \sim \mathcal{O}[(\sigma
N)^{1/4}]$ at horizon exit. Therefore, from Eqs.~(\ref{eq:tension}),
(\ref{eq:tension_dless}) and (\ref{eq:mass}) one finds for $|h| \lesssim 1$,
\begin{align}
m \simeq \frac{(5 \sqrt{\pi}\,A_s^{cmb})^{2/3}}{N_\star^{5/6}}\, M_5 \simeq
10^{-4}\,M_5\,.
\end{align}
Moreover, as in the SC case, large values of $\vert h \vert$ imply smaller
values of $m$ due to an extra suppression $\mathcal{O}(1/|h|)$.

After inflation ceases, the inflaton field starts to oscillate near the minimum
of the effective potential, gradually  producing a large number of particles,
which interact with each other and come to a state of thermal equilibrium at
some temperature $T_{\textrm{reh}}$, the reheating temperature. While we do not
wish to commit ourselves to any specific reheating model, thus keeping our
discussion as general as possible, we can make a rough estimate of the
reheating temperature using the relation between $N_\star=N(k_\star)$, the
number of e-folds before the end of inflation when the scale of wavenumber
$k_\star$ crosses the Hubble radius during inflation, i.e. when $k=aH$, and the
energy density at the end of the reheating period, $\rho_{\textrm{reh}}$.
Although such a relation crucially depends on the entire history of the
universe, a plausible estimate can be obtained under simple assumptions about
the sequence of epochs which follow inflation. We also recall that in the
high-energy regime of brane cosmology the expansion laws during the matter and
radiation dominated eras are slower than in standard cosmology. Nevertheless,
the behavior of the densities is unchanged with respect to the scale factor.

Using the slow-roll approximation during inflation one can
write~\cite{Liddle:2003as}
\begin{align}
N_\star =  &\, \ln \dfrac{k_\star^{-1}}{3000\, h^{-1} \mbox{Mpc}} +
\dfrac{1}{12}\ln \dfrac{\rho_{\textrm{reh}}}{\rho_{\textrm{end}}} +
\dfrac{1}{4}
\ln \dfrac{\rho_{\textrm{eq}}}{\rho_{\textrm{end}}}\nonumber\\
&+\ln \dfrac{H_\star}{H_{\textrm{eq}}} + \ln 219\, \Omega_m\, h\,,
\label{eq:reh}
\end{align}
where $H_{\textrm{eq}}$ and $H_\star=H(k_\star)$ are the values of the Hubble
radius at the matter-radiation equality epoch and at the scale $k_\star$,
respectively; $\rho_{\textrm{end}}$ and $\rho_{\textrm{eq}}$ are the values of
the energy density at the end of inflation and at equilibrium; $\Omega_m$ is
the fractional matter energy density at present. We have
\begin{align}
H_{\textrm{eq}}&=5.25\times 10^6 h^3\, \Omega_m^2\, H_0~,\\
H_0& =8.77\times 10^{-61} h\, M_P~,\\
\rho_{\textrm{reh}}&=\dfrac{\pi^2}{30}\, g_*\, T_{\textrm{reh}}^4\,,
\end{align}
where  $H_0$ is the present Hubble radius and $g_* \simeq \mathcal{O}(10^2)$ is
the effective number of relativistic degrees of freedom. The CMBR anisotropy
measured by WMAP allows a determination of the fluctuation amplitude at the
scale $k=0.002~\mbox{Mpc}^{-1}$ and we use the WMAP3 central values $h=0.73,
~\Omega_m h^2=0.127$ \cite{Spergel:2006hy}.

We find that, for an inflationary period driven by the power-law potential of
Eq.~(\ref{eq:potphi}), the relation
\begin{align}
T_{\textrm{reh}}^{SC} \simeq (10^7-10^8)\times e^{3(N_\star -55)}~\mbox{GeV}~,
\end{align}
provides a good fit to our numerical results in the SC limit, as we allow
$N_\star$ to vary in the expected range $ 55 \leq N_\star \leq 70$ (see
discussion in Section III). Similarly, in the high-energy regime of BC we get
\begin{align}
T_{\textrm{reh}}^{BC} \sim 10^2\, \dfrac{M_5}{M_P}\, T_{\textrm{reh}}^{SC}~,
\end{align}
which is valid for $M_5 \lesssim 10^{-2} M_P \approx 10^{16}$~GeV. For larger
values of $M_5$, $M_5 > 10^{16}$~GeV, one has $T_{\textrm{reh}}^{BC} \simeq
T_{\textrm{reh}}^{SC}$. We also remark that the variation of $T_{\textrm{reh}}$
as a function of the density fluctuation parameters $n_s$ and $r_s$ is mild in
both the SC and BC regimes.

\section{Conclusions}
\label{sec:conclusion}

In this paper we have explored inflationary solutions for a single-field
polynomial potential, with up to quartic terms in the inflaton field, in light
of the results from the WMAP three-year sky survey. Our analysis is done in the
framework of the Randall-Sundrum II braneworld theory, and we have examined
both the high-energy and standard cosmology regimes. As in the case of SC, the
model displays large-field and small-field inflationary type solutions
compatible with the observational data.

For a potential with a negative mass square term, cases A and B of
Table I, we conclude that small-field inflation is in general
favored by current bounds on $r_s$. We also get the bound
$\widetilde{\lambda} \lesssim 10^{-13}$ on the quartic coupling of
the inflaton potential. The case with positive mass square term
and $\vert h\vert\leq 1$ (large-field solutions), which
corresponds to case C of Table I, is also very much constrained by
the $r_s$ bound and, in particular, this type of solutions  will
be excluded if, as expected, $r_s < 0.2$. In this case we get the
following bounds on the parameters of the inflationary potential:
$\widetilde{\lambda} \lesssim 10^{-12}$ and $|\widetilde{g}|
\lesssim 10^{-6}$.

We have also made an estimate of the reheating temperature for this model and
we have found that, for $M_5 > 10^{16}$~GeV and $N_\star \leq 60$, one has
$T_{\textrm{reh}}^{BC} \simeq T_{\textrm{reh}}^{SC} \lesssim 10^{13}-10^{14}$
GeV, whereas for $M_5 < 10^{16}$~GeV, one gets the relation
$T_{\textrm{reh}}^{BC} \sim 10^2\, T_{\textrm{reh}}^{SC}{M_5}/{M_P}\,$. At this
point, it is worth noting that in supergravity models the presence of the
gravitino could lead to serious cosmological problems unless the reheating
temperature is sufficiently low. In particular, in gravity-mediated
supersymmetry breaking models, the gravitino mass is expected in the range
$m_{3/2} \sim \mathcal{O}(10^{2-4})$~GeV. Such a gravitino is most likely
unstable and its decays could destroy the light elements synthesized during
BBN. On the other hand, in gauge-mediated supersymmetry breaking models, the
gravitino can be the lightest supersymmetric particle, and thus stable, with a
mass $m_{3/2} \lesssim \mathcal{O}(10)$~GeV. In this case, its contribution to
the present cosmic density can be excessive. In either case, this leads to
stringent constraints on $T_{\textrm{reh}}$ after
inflation~\cite{Kawasaki:2006hm}. In the braneworld scenario, one expects such
bounds to translate into additional constraints on the 5D fundamental Planck
mass $M_5$.

Finally, let us comment on the running of the scalar index
$\alpha_s$. One should notice that it is not possible to obtain
large values for $\alpha_s$, either in SC or the high-energy BC
scenarios, for polynomial potentials of the form given in
Eq.~(\ref{eq:potphi}). However, this is still consistent with
observation, since the evidence for running is still weak and it
can evaporate as more data becomes available. In fact, the
addition of Lyman-$\alpha$ forest data reduces the need for a
negative value of
$\alpha_s$~\cite{Seljak:2004xh_plus,Lewis:2006ma,Seljak:2006bg}.
Recently Easther $\&$ Peiris~\cite{Easther:2006tv} have analyzed
the implications of a large running spectral index for
single-field slow-roll inflation in SC, using the inflationary
flow equations~\cite{Hoffman:2000ue_plus,Liddle:2003py} and
retaining all terms in $\alpha_s$ up to quadratic order in the
slow-roll parameters. They found that, for all parameter choices
consistent with a large negative running, inflation lasts less
than 30 e-folds after CMBR scales leave the horizon. They
concluded that a definitive observation of a large negative
running would imply that inflationary stage requires multiple
fields or the breakdown of slow-roll. The underlying conclusions
of their work can presumably be carried over to the BC case, as
the flow equations are quite insensitive to the expansion
dynamics~\cite{Ramirez:2004fb}. However, one should notice, as
shown in~\cite{Liddle:2003py}, that the flow equation approach
does not directly incorporate inflationary dynamics. In fact, one
can determine analytically the set of inflationary potentials
which correspond to solutions of the truncated flow equations.
Hence, conclusions obtained from an analysis based on the flow
equations should be considered with some care, since it is
certainly possible to find potential shapes that are not
represented by the set of inflationary potentials for which the
formalism is valid. An example of a model where it is possible to
obtain a considerably large running was explored in
Ref.~\cite{Ballesteros:2005eg}.

\begin{acknowledgments}
The work of R.G.F. and M.C.B. was supported by Funda\c c\~ao para a Ci\^encia e
a Tecnologia (FCT, Portugal) under the grant SFRH/BPD/1549/2000 and the project
POCI/FIS/56093/2004, respectively. This work was also supported by FCT through
the projects POCI/FIS/56093/2004 and PDCT/FP/FNU/50250/2003.
\end{acknowledgments}


\end{document}